\documentclass[prd,twocolumn,nofootinbib,floatfix,preprintnumbers,letterpaper]{revtex4}
\usepackage{graphicx}
\input{epsf}
\usepackage{epsf}
\usepackage{graphicx,epsfig}
\usepackage{bm}
\usepackage{latexsym,amssymb,amsmath,float}

\def\rd{{\rm d}}

\def\bs{\begin{split}}
\def\es{\end{split}}

\def\5{\overline 5}

\renewcommand{\(}{\left(} 
\renewcommand{\)}{\right)} 
\renewcommand{\[}{\left[} 
\renewcommand{\]}{\right]}

\newcommand{\rf}[1]{(\ref{#1})}

\newcommand{\half}{\frac{1}{2}} 
\newcommand{\beq}[1]{\begin{equation}\label{#1}}
\newcommand{\eeq}{\end{equation}}
\newcommand{\bea}[1]{\begin{eqnarray}\label{#1}}
\newcommand{\eea}{\end{eqnarray}}

\def\be{\begin{equation}}
\def\ee{\end{equation}}
\def\ba{\begin{eqnarray}}
\def\ea{\end{eqnarray}}

\begin{document}

\date{\today}

\title{A new perspective on the relation between dark energy perturbations\\ 
       and the late-time ISW effect}

\author{James B. Dent, Sourish Dutta, and Thomas J. Weiler}
\affiliation{Department of Physics and Astronomy, Vanderbilt University,
Nashville, TN  ~~37235}

\begin{abstract}
The effect of quintessence perturbations on the ISW effect is studied for a mixed dynamical scalar field 
dark energy (DDE) and pressureless perfect fluid dark matter. A new and general methodology is developed to track the growth of the perturbations, which uses only the equation of state (EoS) parameter $w_{\rm DDE} (z) \equiv p_{\rm DDE}/\rho_{\rm DDE}$ of the scalar field DDE, and the initial values of the
the relative entropy perturbation (between the matter and DDE) and the intrinsic entropy perturbation of the scalar field DDE as inputs. We also derive a relation between the rest frame sound speed $\hat{c}_{s,{\rm DDE}}^2$ of an arbitrary DDE component and its EoS $w_{\rm DDE} (z)$. 
We show that the ISW signal differs from that expected in a $\Lambda$CDM cosmology by 
as much as 
+20\% to -80\% for parameterizations of $w_{\rm DDE}$  
consistent with SNIa data, and about $\pm$ 20\%  for parameterizations of $w_{\rm DDE}$ consistent with
SNIa+CMB+BAO data, at $95\%$ confidence. 
Our results indicate that, at least in principle, the ISW effect 
can be used to phenomenologically distinguish a cosmological constant from DDE. 
\end{abstract}

\maketitle

\section{Introduction}
\label{sec:introduction}

It has been known for almost a decade that the universe is accelerating~\cite{Knop,Riess1,Wood-Vasey,Davis}, 
and considerable evidence has accumulated which indicates that the acceleration is due to a negative energy 
component constituting $70\%$ of the energy density of the Universe~\cite{Komatsu:2008hk,Tegmark:2006az, Percival:2006gs}. 
Determining the nature of this dark energy has become a central challenge of cosmology. 
 
A first step in this direction would be to determine whether the dark energy is sourced by a cosmological constant 
(for reviews, see~\cite{padmanabhan}) or a dynamical field (for reviews, see~\cite{copeland}). In terms of physical properties, 
a dynamical field may be distinguished from a cosmological constant by a time varying equation of state (EoS)  
$w(z) \equiv p(z)/\rho(z)$, 
and a non-zero (squared) sound speed $c^2 (z)\equiv \delta p(z)/\delta\rho (z)$. 
As we discuss below, both of these quantities could result in an observable signature different from a cosmological constant. 

Both $p$ and $\rho$ are functions of the field (scalar in the case of quint- or k-essence) 
associated with the dynamical dark energy, and in principle can be fully determined by solving the equations of motion of this field.  
Alternatively, one can parameterize $w(z)$ as a function of red-shift without reference to an explicit form 
of the Lagrangian for the field.  A representative list of parameterizations is given in~\cite{Lazkoz:2005sp}, 
and scalar field reconstruction from these parameterizations is given in~\cite{liguozhang, liholzcooray}.  

The ``sound speed'' of a generic fluid  is defined as $c^2 \equiv \delta p_{\rm DDE}/\delta\rho_{\rm DDE}$.
The name ``sound speed'' is a misnomer for a non-thermal component, and so $c^2 (z)$ can be thought of as 
shorthand for $\delta p_{\rm DDE}/\delta\rho_{\rm DDE}$.\footnote
{The thermodynamic squared sound speed is $c_s^2=(\delta p/\delta\rho)_{\rm S}$.
For an adiabatic system, by definition the entropy~S is constant, and so the adiabatic sound speed $c_a^2=\delta p/\delta\rho$.
For non-adiabatic systems one needs to take into account the non-adiabatic part of the pressure perturbation $\delta p_{\rm nad}$, leading to a non-adiabatic component to the sound speed. 
}  
The sound speed has been a less studied source of insight into the 
nature of dynamical dark energy than has the EoS, 
but its effects have been investigated recently in~\cite{dedeocaldwellsteinhardt,beandore,huscranton}.  
Sound speed varies widely among different DDE models.
For example, quintessence, with its canonical kinetic term, has a constant sound speed equal to unity (in the rest-frame of quintessence). 
On the other hand, k-essence has a non-canonical kinetic term, leading to a $z$-dependent sound speed.
%
%
Other DDE candidates, e.g., dilatons, Chaplygin gas, phantoms, and tachyons, each have their own 
unique attributes determining a characteristic sound speed.
As we show later, the evolution of the sound speed depends on the EoS parameter of the dark energy component, 
as well as the dark energy's intrinsic and relative (to the matter component) entropy perturbations.

The dynamical evolution of the field can affect a number of physical observables, 
including the spectrum and growth of large scale structure, weak gravitational lensing, 
SNIa~apparent luminosities, and CMB anisotropies.   
These avenues have been explored in various works ~\cite{beandore,huscranton,cooray,garrigapogosianvachaspati,pogosian}. 
One particular manifestation occurs in the late-time Integrated Sachs-Wolfe (ISW) effect, 
which measures the evolution of the gravitational potential as the Universe enters 
a phase of dark energy domination at $z\alt 2$.
This effect is only significant on large scales (low multipoles), 
since small-scale fluctuations in the gravitational potential smooth out along the line of sight.  
And it is only significant at late times since potentials evolve the same as the background during matter domination.
The late-time ISW effect has been detected in cross-correlations between CMB temperature anisotropies 
and surveys of large scale 
structure~\cite{Boughn:1997vs,Fosalba:2003iy,Gaztanaga:2004sk,Vielva:2004zg,Pietrobon:2006gh,
Giannantonio:2006du,McEwen:2006my,Rassat:2006kq,Zaldarriaga:1998te,Hu:2001kj,Song:2002sg,
Kaplinghat:2003bh,Hu:2001fb,Hu:2001tn,Ho:2008bz,Granett:2008ju}. 

Recent work on the clustering properties of scalar field DDE have indicated that, in the context of Einstein's 
general relativity, scalar field dark energy perturbations are likely to be anti-correlated to 
matter perturbations in the linear regime~\cite{wellerlewis,Dutta:2006pn,motashawsilk}. 
In~\cite{Dutta:2006pn} it has been shown that even if the scalar field is initially homogeneous,  
it eventually acquires a perturbation anti-correlated to the matter perturbation as a result of gravitational coupling. 

At low redshifts, 
very large scale perturbations (on the scale of the horizon) are linear and are responsible for the 
late-time ISW signal in the CMB as described above.  For a $\Lambda$CDM 
cosmology, the ISW peak in the CMB is purely the effect of matter perturbations, of course. 
The question we ask in this paper is how much the ISW signal from scalar field DDE linear perturbations 
differs from that due to the $\Lambda$CDM.  

To answer this question we study the role played by both the EoS parameter 
and the (relative and intrinsic) entropy perturbations of the dark energy component in the evolution of the Newtonian potential. 
We compare effects to the case of a cosmological constant. Our approach is different from 
previous treatments in that we apply the equations of linear perturbation theory to the 
case of a generic quintessence component characterized by a parameterized EOS $w(z;w_0,w_1)$ rather than 
by a Lagrangian, allowing for considerable generality in the inclusion of a quintessence component of the Universe. Our gauge-invariant approach also tracks the evolution of entropy perturbations in a consistent manner. 

The outline of this paper is as follows: in the next section (II) we describe our mathematical 
approach and demonstrate how it connects with previous treatments. We summarize two popular 
parameterizations of $w_{\rm DDE} (z)$ evolution, and we present a new, ``rapid transition'' parameterization, and then go on to discuss initial conditions.
Our results can be found in section~\ref{Results}, followed by conclusions in~\ref{Conclusions}. 

\section{Modelling the system}

\subsection{Matter and metric perturbations}
We begin with a very general perturbed metric.
In this Section, we follow the notation found in~\cite{copeland,Hwangtachyon,Hwang05} to write
\begin{eqnarray}
\hspace*{-0.2em}\rd s^2 &=& - (1+2A)\rd t^2 +
2a\partial_iB \rd x^i\rd t
\nonumber\\
\hspace*{-0.2em}&& +a^2\left[ (1+2\psi)\delta_{ij} + 2\partial_{ij}E 
\right] \rd x^i \rd x^j\,,
\end{eqnarray}
where $A$,$B$,$\psi$ and $E$ represent metric perturbations and $a$ represents the cosmic scale factor. 

We work with a generic cosmic mixture of matter+scalar field DDE, characterized by a total pressure $p$, 
total density $\rho$, averaged velocity potential $v$, total EoS parameter $w=p/\rho$ and total sound speed $c^2=\delta p/\delta\rho$.  
We adopt the convention that variables and parameters 
characterizing the total fluid are presented without subscripts, whereas variables and parameters
characterizing single components of the fluid are presented with component-identifying subscripts.
We assume that the matter component is a perfect fluid. Several possibilities exist for the scalar field DDE component - in our approach we simply characterize its equation of state parameter as $w_{\rm DDE}\(z;w_0,w_1\)$, 
where ${w_0,w_1}$ are arbitrary parameters, discussed in Section~\rf{subsec:DDE_EoS}. 

We now parallel the formalism in~\cite{copeland}, presented in the 
longitudinal gauge~\cite{Mukhanov:1990me,mabertschinger}.
This gauge choice corresponds 
to a transformation to a frame such that $B=E=0$. 
In this gauge, the physical gauge-invariant variables which characterize 
the metric perturbations become:
\begin{eqnarray}
\label{defPhi2}
\Phi &\equiv& A - \frac{{\rm d}}
{{\rm d}t} \left[ a^2(\dot{E}+B/a)\right] \rightarrow A\,,\\
\label{defPsi2}
\Psi &\equiv& -\psi + a^2 H (\dot{E}+B/a) \rightarrow -\psi\,.
\end{eqnarray}
The energy-momentum tensor can be decomposed as 
\begin{eqnarray}
& & T_0^0=-(\rho+\delta \rho)\,,\quad 
T^0_{\alpha}=-(\rho+p) v_{,\alpha}\,, \nonumber \\
& & T^{\alpha}_{\beta}=(p+\delta p) 
\delta^{\alpha}_{\beta}+\Pi^{\alpha}_{\beta}\,,
\end{eqnarray}
where $\Pi^{\alpha}_{\beta}$ is a tracefree anisotropic stress.
Henceforth, we assume that the anisotropic stress $\Pi^{\alpha}_{\beta}$ is zero.

The perturbed Einstein equations yield, at linear order,
\begin{eqnarray}
\label{pereq1}
-\Phi+\Psi&=&0\\
\label{pereq2}
 -\frac{\Delta}{a^2}\Phi+3H^2\Phi + 3H\dot{\Phi}&=&-4\pi G \delta \rho \\
\label{pereq3}
H\Phi+\dot{\Phi}&=&4\pi G a(\rho+p)v\\
\label{pereq4}
3\ddot{\Phi} + 9H\dot{\Phi}\qquad\qquad\qquad&& \nonumber\\
  +(6\dot{H} + 6H^2 +\frac{\Delta}{a^2})\Phi&=&4\pi G (\delta \rho+3\delta p) \\
\label{pereq5}
\delta \dot{\rho}+3H (\delta \rho+\delta p)&=&(\rho+p) \left(3\dot{\Phi}\right.\nonumber\\
&&+\left.\frac{\Delta}{a}v\right) \\
\label{pereq6}
\frac{[a^4(\rho+p)v]^{\bullet}}{a^4(\rho+p)}&=&\frac{1}{a} \left( A+\frac{\delta p}{\rho+p}
\right)
\end{eqnarray}
where a dot, bold or otherwise, denotes a derivative with respect to coordinate time t, 
$H \equiv \dot{a}/a$, and any quantity preceded by $\delta$ denotes a perturbation in that quantity.  
We have also used the relation between $\Phi$ and $\Psi$ in Eq.~(\ref{pereq1}) in the subsequent equations.  
This relation simply reflects our assumption of no anisotropic stress.  
The resulting anisotropic stress-free line element in the longitudinal gauge is simply
\beq{simpleLE}
ds^2=-(1+2\Phi)dt^2 +a^2(1-2\Phi)d{\vec x}^2\,.
\eeq
From this point on we will use the Newtonian potential $\Phi$ to characterize the metric perturbation.

Since we work with linear perturbation theory, it is extremely convenient to  transform to Fourier space.
The Fourier modes evolve independently. For the remainder of this paper, we work with individual modes in Fourier space.
For convenience we suppress the $k$-subscripts, and ask the reader to keep this in mind.

We first express the total matter overdensity ($\delta\equiv\delta\rho/\rho$) 
and the useful combination $\Theta\equiv\(c^2-w\)\delta$ 
entirely in terms of $\Phi$ and the background variables 
using Eqs.~(\ref{pereq2}) and (\ref{pereq4}), as follows:
\begin{eqnarray}
\label{delta}
& & \delta = - \left(\frac{2k^2}{3a^2H^2}\right)\Phi -2\Phi -2\frac{\dot{\Phi}}{H}\\
\label{deldub}
& & \Theta = \frac{2}{3H^2}[ \ddot{\Phi} + H(4 + 3w)\dot{\Phi} + w\frac{k^2}{a^2}\Phi]
\end{eqnarray}

As is sometimes done (e.g.~\cite{mabertschinger,beandore,wellerlewis}), 
we can re-express the system in terms of the matter density contrast $\delta$ and the velocity potential $v$ 
(a scalar field whose gradient is the true velocity) 
as follows:
\ba
\label{delta-dot}
\dot{\delta}&=&-3H\Theta +3(1+w)\dot{\Phi}-(1+w)\frac{k^2}{a}v\\
\label{v-dot}
\dot{v}&=&-vH\(1-3w\)-\frac{\dot{w}}{1+w}v\nonumber\\
&&\quad+\frac{1}{a}\[\Phi+\frac{w}{1+w}\delta+\frac{\Theta}{1+w}\]
\ea

The growth of the matter overdensity can be shown to be governed by the second order equation: 
\be
\label{deltadoubledot}
\begin{split}
\ddot{\delta} + \dot{\delta}(2-3w)H + \frac{k^2}{a}w\delta + \frac{k^2}{a^2}(1+w)\Phi = \\ 
 3(1+w)[\ddot{\Phi} + \dot{\Phi}(2-3w)H] + 3\dot{w}\dot{\Phi} \\
 -3H\dot{\Theta}+ \Theta [ -\frac{k^2}{a^2}+\frac{3H^2}{2}(1 + 9w) ]
\end{split}
\ee
 Eq.~(\ref{deltadoubledot}) can be deduced
 from Eq.~(\ref{pereq6}) after substituting $v$ from Eq.~(\ref{pereq3}), $\delta$ from Eq.~(\ref{delta}), 
and using the time derivative of Eq.~(\ref{pereq5}).  
This equation fully characterizes evolution in terms of the metric and its perturbation $\Phi$,
the overdensity $\delta$, the equation of state parameter $w$, the sound speed $c^2$, and the expansion rate $H$. 
It can be verified that these equations are equivalent to Eq.~(30) in~\cite{mabertschinger}.

The pressure perturbation can be decomposed into an adiabatic and non-adiabatic component as follows:
\be
\label{p0}
\delta p=\delta p_{\rm nad}+c_{a}^2\delta\rho
\ee
where $\delta p_{\rm nad}$ is the non-adiabatic pressure perturbation and $c_{a}^2\equiv \dot{\rho}/\dot{p}$ 
is the adiabatic sound speed of the total fluid. 

Finally, following \cite{Bartolo:2003ad}, we introduce two gauge-invariant entropy perturbation variables. 
The relative entropy perturbation between the matter and the scalar field DDE is denoted by the variable $S$:
\be
\label{S}
S\(z\)\equiv\frac{3H(1+w_{\rm DDE})\Omega_m}{1+w}\(\frac{\delta\rho_{\rm DDE}}{\dot{\rho}_{\rm DDE}} - \frac{\delta\rho_m}{\dot{\rho_m}}   \)
\ee
The intrinsic entropy perturbation of the scalar field DDE is denoted by the variable $\Gamma$:
\be
\label{T}
\Gamma\(z\)\equiv\frac{3H(1+w_{\rm DDE})c_{a,DDE}^{2}}{1-c_{a,DDE}^2}
   \(\frac{\delta\rho_{\rm DDE}}{\dot{\rho}_{\rm DDE}} - \frac{\delta p_{\rm DDE}}{\dot{p}_{\rm DDE}}   \)
\ee

In terms of these entropy perturbations, the non-adiabatic pressure perturbation can be written as:
\ba
\label{p1}
\delta p_{\rm nad} &=& \Omega_{\rm DDE}\rho\[\(- c_{a,{\rm DDE}}^2\)S\right. \\\nonumber &+& \left.\(1-c_{a,{\rm DDE}}^2\)\Gamma\]
\ea

\subsection{Connection with previous work}
The system of equations~(\ref{pereq1}-\ref{pereq6}) 
and the equations~(\ref{delta}-\ref{deltadoubledot}) derivative from them 
are perfectly general, following directly from the Einstein equations.
We now digress to show how what we have done to this point connects with  previous work.

Refs.~\cite{beandore} and~\cite{wellerlewis} adopt the approach of starting with a set of equations 
similar to Eq.~(\ref{delta-dot}) and Eq.~(\ref{v-dot}), 
and then proceeding with simplifying assumptions regarding the behavior of the component(s) of the matter fluid. 

In Eq.~(\ref{deltadoubledot}), if we assume (as is done in~\cite{beandore}) 
that $c^2 = w$ and $d(c^2)/dt = 0$ (these assumptions are equivalent to setting $\dot{w}$, 
$\Theta$ and $\dot{\Theta}$ to zero), we obtain
\be
\label{deltadoubledot2}
\begin{split}
\ddot{\delta} + \dot{\delta}(2-3w)H + \frac{k^2}{a}w\delta + \frac{k^2}{a^2}(1+w)\Phi \qquad \\
=3(1+w)[\ddot{\Phi} + \dot{\Phi}(2-3w)H]\,,
\end{split}
\ee
which coincides with Eq.~(375) in~\cite{copeland}. 
In addition, ignoring the last two terms of Eq.~\rf{delta} 
so that $\Phi$ is expressed by the Poisson equation, and assuming that the evolution 
is largely matter dominated ($w\simeq0$), one obtains the more familiar form

\ba
\label{growth}
\ddot{\delta} + 2H\dot{\delta}  - 4\pi G\rho\delta = 0 \,.
\ea
This equation is Eq.~(66) of~\cite{padmanabhan} and the (unnumbered) starting equation in~\cite{cooray}. 

Finally, assuming a late time DDE dominance, defining a growth factor $g(a) \equiv \delta(a)/a$ in the usual way, 
and rewriting in terms of conformal time and its derivatives (denoted by primes),
this equation~\rf{growth} becomes
\be
\label{cooray eqn}
\begin{split}
g'' + \left(\frac{5}{2} -\frac{3w(a)\Omega_{\rm DDE}(a)}{2}\right)g' \qquad\qquad\qquad\\
+ \frac{3}{2}(1-w(a))\Omega_{\rm DDE}(a) = 0 \,.
\end{split}
\ee
This is just Eq.~(1) of~\cite{cooray}.

In Ref.~\cite{huscranton}, the authors use Eq.~\rf{cooray eqn} with the metric perturbation $\Phi$ instead of $g$. 
Clearly, this equation is valid \textbf{in the longitudinal gauge} only if firstly, one assumes that $\Phi$ and $\delta$ are connected through the Poisson equation, i.e.,
$\Phi\simeq 4\pi G a^3\delta/k^2 \sim 4\pi G a^3 \rho g/k^2$
and secondly, the Universe is largely matter dominated, which implies that $\rho\sim a^{-3}$. 
Together, these imply that $\Phi\propto g$, making Eq.~\rf{cooray eqn} valid for the metric perturbation as well.

On a side note, Eq.~(\ref{growth}),  is commonly known as the ``growth equation''. 
While this equation is exactly correct in the synchronous gauge, it is only approximate in the longitudinal gauge. 
A detailed study examining the accuracy of this equation on different scales in the longitudinal gauge was performed 
in \cite{dentdutta}, where it was found that it can be surprisingly inaccurate on scales larger than $\sim 0.1h$ Mpc$^{-1}$. 
The chief cause for the breakdown of this equation on large scales 
was shown to be the replacement of Eq.~\rf{delta} by the Poisson equation, and a modified growth equation was 
proposed as a better approximation for calculation performed in the longitudinal gauge.

\subsection{Evolution equation for metric perturbation {\boldmath $\Phi$}}

In this paper we avoid making assumptions regarding the specific nature or behavior of the scalar field DDE component. 
Instead, we express the evolution of $\Phi$ in terms of the scalar field DDE EoS $w_{\rm DDE}\(z;w_0,w_1\)$, 
and the two independent entropy perturbations $S$ and $\Gamma$. 
Switching to redshift~$z$ as the time variable,
we obtain from Eqs.~(\ref{delta}), (\ref{deldub}) and (\ref{p1}) and Eqs.~(20), (33), and (34) of \cite{Bartolo:2003ad},\\ 
{\sl our central result}: 
\begin{widetext}
\ba
\label{phiz}
\frac{d^2\Phi}{dz^2} &=&- \frac{1}{1+z}\frac{d\Phi}{dz}\left(\frac{3}{2}w - 3c_{a}^2 - \frac{3}{2}\right) 
-\Phi\left(\frac{c_{a}^2 k^2}{H(z)^2} - 3\frac{(c_{a}^2 -w)}{(1+z)^2}\right) \nonumber\\
&&+\frac{3}{2}\frac{\Omega_{\rm DDE}}{\(1+z\)^2}\[-c_{a,{\rm DDE}}^2 S+\(1-c_{a,{\rm DDE}}^2\)\Gamma\]\\
\frac{dS}{dz}&=&-\frac{1}{1+z}\[ \(3w_{\rm DDE}-\frac{3\Omega_{m}c_{a,{\rm DDE}}^2}{1+w}\)S
   +3\frac{3\Omega_{m}\(1-c_{a,{\rm DDE}}^2\)}{1+w}\Gamma \right.\nonumber\\
\label{Sz}&&\left. \frac{k^2 \(1+z\)^2}{H^2}\(\frac{1}{3}S+\frac{1}{3}\Gamma\)
   +\frac{ k^4 \(1+z\)^4}{H^4}\(\frac{2}{9}\frac{\(1+w_{\rm DDE}\)}{\(1+w\)}\Phi\)\]\\
\frac{d\Gamma}{dz}&=&-\frac{1}{1+z}\[-\frac{3}{2}\(1+w\)S+3\(w_{\rm DDE}-\frac{1+w}{2}\)\Gamma\right.\nonumber\\
\label{Gammaz}&&\left. \frac{k^2 \(1+z\)^2}{H^2}\( -\(1+w_{\rm DDE}\){\cal R}-\frac{1}{3}S-\frac{1}{3}\Gamma\) 
   +\frac{ k^4 \(1+z\)^4}{H^4}\(-\frac{2}{9}\frac{\(1+w_{\text{DDE}}\)}{\(1+w\)}\Phi\)      \] \,.
\ea
\end{widetext}

The different quantities appearing in this system of equations are discussed below. \textbf{ It is important to note that apart from the evolved variables ($\Phi$,$S$,$\Gamma$) and $\cal R$ (which is a combination of $\Phi$ and $d\Phi/dz$) all the variables appearing in the above equations can be expressed as functions of the parameterized scalar field DDE EoS $\mathbf{w_{\textbf{DDE}}(z)}$}, as we show below. For convenience, these variables are listed in Table \ref{vars}.

The quantity $\cal R$, appearing in the final line of Eq.~\ref{Gammaz}, 
is the gauge-invariant comoving curvature perturbation defined as
\be
\label{scriptR}
{\cal R}\equiv\Phi+\frac{2}{3\(1+w\)}\[\Phi-\(1+z\)\frac{d\Phi}{dz} \]
\ee

Note that the dynamical Hubble parameter $H(z)$ appearing in Eq.~(\ref{phiz}) 
is a known function of $z$: 
\ba
\label{Hz}
\frac{H^2(z)}{H^2 (0)}&=& 
\left[ 
\Omega_m (0)\,(1+z)^3 \right.\\\nonumber
&+& \left.\Omega_{\rm DDE} (0)\,e^{3\int dz\,\frac{1 + w_{\rm DDE}(z)}{1 + z}}
\right] \,.
\ea
$H(0)$ is the Hubble parameter today, measured to be $72\pm5$~km/s/Mpc~\cite{keyproject}.
The values $\Omega_m (0)$ and $\Omega_{\rm DDE} (0)$ are related by $\Omega_m (0)+\Omega_{\rm DDE} (0)-1\propto$
the curvature of the Universe.  Since the first Doppler peak of the CMB offers strong evidence
that the Universe is flat, we make the standard inference that $\Omega_m (0)+\Omega_{\rm DDE} (0)=1$.

Moreover, the total EoS parameter $w(z)$ is determined by the field EoS, $w_{\rm DDE}$, as follows.
In general, $w$ is related to the individual EoS parameters $w_{i}$ 
of the individual components, each with density parameter $\Omega_{i}\equiv\rho_i/\sum\rho_i$, as 
\begin{equation}
\label{wtotal}
	w(z)=\sum\Omega_{i}(z)\,w_{i}(z)\,.
\end{equation}
Hence, for a mixture of perfect fluid matter and scalar field DDE, the total EoS 
parameter $w$ is just 
\begin{eqnarray}
\label{wz}
w(z) &=& \Omega_{\rm DDE}(z)\,w_{\rm DDE}(z)\nonumber\\
     &=& \Omega_{\rm DDE} (0)\,\left[ 
    \frac{H_{0}^2}{H^2 (z)}\,e^{3\int_0^z {dz\left( {\frac{{1 + w_{\rm DDE}  }}{{1 + z}}}\right)}}
    \right]\,w_{\rm DDE} (z)\,. \nonumber\\
\end{eqnarray}

The adiabatic sound speed of the total fluid $c_{a}^2(z)$ is computed to be 
\begin{equation}
\label{csa}
c_{a}^2(z)\equiv\frac{\dot{p}}{\dot{\rho}} = w+\frac{1+z}{3\left(1+w\right)}\frac{dw}{dz}\,,
\end{equation}
leading to 
\begin{equation}
\label{csa_full}
c_{a}^2 \left( z \right) = \frac{w}{1+w}\left[\left(1+w_{\rm DDE}\right)
   +\frac{\left(1+z\right)}{3}\frac{w_{\rm DDE}'}{w_{\rm DDE}}\right]\,.
\end{equation}
The adiabatic sound speed of the DDE component is given by 
\be
\label{ca_sq_dde}
c_{a,{\rm DDE}}^2=w_{\rm DDE}+\frac{1}{3}\frac{w_{\rm DDE}'(z)\(1+z\)}{1+w_{\rm DDE}}
\ee

\begin{table*}
		\centering
		\begin{tabular}{|l|p{4.5cm}|l|}
		\hline
		\textbf{Symbol}       & \textbf{Name}               & \textbf{Defined in }\\\hline
		$\Phi$ 	              & metric perturbation         & Eq. \ref{simpleLE}        \\\hline
		$S$                   & relative entropy perturbation of the DDE  & Eq. \ref{S}       \\\hline
		$\Gamma$              & intrinsic entropy perturbation of the DDE & Eq. \ref{T}        \\\hline
		$\cal R$              & comoving curvature perturbation & Eq. \ref{scriptR}          \\\hline
		$w$                   & EoS of the total fluid                    & Eqs. \ref{wtotal}, \ref{wz}       \\\hline
		$w_{\rm DDE}$         & parameterized EoS of the DDE & Eqs. \ref{linder_param}, \ref{log_param}, \ref{RTparam}          \\\hline
		$c_{a}^2$             & adiabatic sound speed of the total fluid  & Eq. \ref{csa_full}          \\\hline
		$c_{a,{\rm DDE}}^2$   & adiabatic sound speed of the DDE & Eq. \ref{ca_sq_dde}          \\\hline
		$\hat{c}_{s,{\rm DDE}}^2$   &rest frame sound speed of the DDE & Eq. \ref{cs_hat}          \\\hline
		\end{tabular}
	\caption{List of important variables. }
	\label{vars}
\end{table*}

The system of Eqs.~(\ref{phiz})-(\ref{Gammaz}) is the central result of this paper. 
The evolution of the metric perturbation 
(for each scale $k$) is characterized entirely by a system of  linear, ordinary differential equations 
for the evolution of Newtonian $\Phi(z)$, 
with only the EoS $w(w_{\rm DDE}(z))$, and the initial values of the entropy perturbations 
$S$ and $\Gamma$ as inputs. These equations present a significant advantage of generality in a gauge-invariant formulation. They allow one to track the behavior of different variables such as $\delta$,$v$, and, (as explained further below) the dark energy sound speed for different classes of dark energy models. 

To the best of our knowledge, the coupled set of equations Eqs.~(\ref{phiz})-(\ref{Gammaz}) has not been written down in previous literature.

Finally, we derive a connection between the (gauge-invariant) dark energy sound speed in the rest frame of the dark energy $\hat{c}_{s,{\rm DDE}}^2$ and the EoS of the dark energy. The pressure perturbation of a species ``i'' in a general frame can be related its to the rest-frame speed of sound  as  follows \cite{beandore}:
\begin{eqnarray}
\label{p2}
\delta p_i = \hat{c}_{s,i}^2\delta\rho_i + 3aH(1 + w_i)(\hat{c}_{s,i}^2 - c_{a,i}^2)\rho_i v_i
\end{eqnarray}

Using the definition of $\Gamma$ (Eq.~(\ref{T})), one can deduce the following expression for $\hat{c}_{s,{\rm DDE}}^2$ as follows:
\begin{eqnarray}
\label{cs_hat}
\hat{c}_{s,{\rm DDE}}^2 &=& c_{a,{\rm DDE}}^2\\\nonumber
                        &+&\frac{\(1-c_{a,{\rm DDE}}^2\)\Gamma}{\delta_{\rm DDE}+3aH\(1+w_{\rm DDE}\)v_{\rm DDE}}
\end{eqnarray}
Clearly, the evolution of the rest-frame DDE sound-speed is linked to the evolution of the EoS parameter $w_{\rm DDE}(z)$ and the intrinsic entropy perturbation $\Gamma(z)$. It is straightforward to check that the rest frame sound speed of scalar field quintessence is identically unity. 

Note that Eq.~(\ref{cs_hat}) is essentially identical to Eq.~(8) of \cite{Hu:1998kj} which was derived in the context of generalized dark matter (subject to a few trivial differences in convention- Hu works in conformal time and $\Gamma_{\rm Hu}=\Gamma_{\rm us}\(1-c_{a,{\rm DDE}}^2\)$).


\subsection{The late-time ISW effect}
\label{sub:ISW}

The integrated Sachs-Wolfe (ISW) effect is an angular variation in the CMB temperature  
due to a photons encountering a time-varying potential well.
The relation between temperature variation and potential is 
\beq{ISWeqn1}
\left(
\frac{\Delta T \left(\hat{\mathbf n}\right)}{T}
\right)_{\rm ISW}
   = 2\int_{\eta_{r}}^{\eta_0} d\eta\;e^{-\tau\left(\eta\right)}\; 
     \frac{\partial\Phi}{\partial\eta}\left[\left(\eta_0 - \eta\right)\hat{\mathbf n},\eta\right]\,,
\eeq
where $\tau(\eta)$ is the optical depth (also called the photon opacity), 
and the integration is along the photon's trajectory from conformal time $\eta_{r}$ at recombination 
to the present conformal time $\eta_{0}$.
In this work, it seems safe to ignore the small photon opacity $\tau$.
Then, integration of Eq.~\rf{ISWeqn1} is trivial, and the resulting ISW relation is simply 
\beq{ISWeqn2}
\left(
\frac{\Delta T \left(\hat{\mathbf n}\right)}{T}
\right)_{\rm ISW}
   = 2\, \left( \Phi[\eta_0] - \Phi[(\eta_0-\eta_r)\hat{\mathbf n},\eta_r ] \right) \,.
\eeq

We have arrived at the ISW relation in Fourier space, i.e., as a 
relation between $\Delta T/T$ and $\Delta\Phi$ valid for each Fourier mode.  
This relation could be inverse Fourier transformed to configuration space,
but there is no clear theoretical advantage in doing so.

Thus, to calculate the ISW effect for a given model, we are left to evolve the Newtonian 
potential $\Phi$ from the time of recombination to the present day, using 
equations Eqs.~(\ref{phiz})-(\ref{Gammaz}). 
As inputs, we need the function $w_{\rm DDE}(z)$ 
and the initial conditions for the variables $\Phi(z)$ and $\dot{\Phi}$,
and for the entropy perturbations $S(z)$ and $\Gamma(z)$.  
As we have mentioned, the evolution of a potential well does not differ from that of a matter-dominated background.
However, we must integrate from a much earlier time to properly include the evolution of the entropy 
perturbations $S$ and $\Gamma$.

\subsection{Parameterizations of the DDE EoS}
\label{subsec:DDE_EoS}
To solve Eq.~(\ref{phiz}),
we adopt a phenomenological approach and parameterize the input function $w_{\rm DDE}(z)$, 
rather than derive it from explicit forms of the matter action. 
Several parameterizations have been suggested for the scalar field DDE EoS parameter $w_{\rm DDE}(z)$.  
Many of these parameterizations have been shown to faithfully mimic the $w$-behavior of well-known DDE 
models~\cite{Corasaniti:2002vg}. Such parameterizations are ideally suited for our formalism, 
as they allow for a simple deduction of the evolution of $\Phi$ for entire classes of models. 
In this paper we choose the following two commonly used forms which are suitably well behaved for large redshifts:

\begin{itemize}
\item The Chevalier-Polarski-Linder (CPL) parameterization~\cite{Chevallier:2000qy, Linder:2002et}
\begin{equation}
	\label{linder_param}
	w_{\rm DDE}(z)=w_{0}+w_{1}\,\frac{z}{(1+z)}
\end{equation}
 
\item The Logarithmic parameterization~\cite{efstathiou}:
\begin{equation}
	\label{log_param}
	 w_{\rm DDE}(z)=w_0+w_{1}\ln\left(1+z\right) 
\end{equation}
\end{itemize}
In these two parameterizations, $w_0$ is today's value $w(0)$.

Several quintessence models have been proposed where the DDE EoS evolves  
from a steady early value to its present value in a rapid transition. 
The EoS in these cases is usually represented by a sigmoidal function~\cite{Corasaniti:2002vg,Bassett:2002qu} 
with four or five parameters. These parameters typically determine the ``final'' value  $w(z=0)$, 
the ``early'' value, the point at which the transition occurs, and the rapidity with which the transition occurs.   
As a representative of this class of models, we choose a much simplified $2$-parameter version:
\begin{itemize}
\item The Rapid Transition (RT) parameterization:
\be
\label{RTparam}
w_{\rm DDE}(z)=\frac{w_0}{1+\(w_{1}z \)^2}\,.
\ee
\end{itemize}
In this parameterization, $w_0$ is again today's value $w(0)$, 
and $|w_1|$ governs both the location in $z$ of the transition 
and the sharpness of the transition. 
The evolution from initial $w=0$ to final $w_0$ is half complete at 
$z_{\half}\equiv 1/|w_1|$.  The slope of $w(z)$ at the half-way point is 
$w'(z_\half)=-w_0\,|w_1|/2=-w_0/2z_\half$.  
Thus, a small $w_1$, say $|w_1|<1$, 
implies an early $z_\half$ and a relatively weak slope (relatively slow transition).
On the other hand, a larger $|w_1|>1$ 
implies a recent $z_\half$ and a relatively large slope (fast transition).
A small $w_1$ (large $z_\half$) gives an EoS differing little from that of the cosmological constant case
where $w=-1$, whereas a large $w_1$ (small $z_\half$) provides a considerable difference.

The functional form for $H^2 (z)$ is given in Eq.~(\ref{Hz}),
and that for $w(z)$ in Eq.~(\ref{wz}).
The same integral exponent appears in both equations.
Each of the three parameterizations we use for $w_{\rm DDE} (z)$ is conspired to allow 
a simple analytic evaluation of the integral exponent in (\ref{Hz}) and (\ref{wz}).

In each of the three parameterizations, 
$w_0$ and $w_1$ are {\sl a priori} arbitrary parameters, 
but later we use the recent SNIa and CMB+BAO data to constrain them.
For each case, we determine numerically the parameter range allowed 
at the $95.4\%\ (2\sigma)$ and $68.3\%\ (1\sigma)$ levels.
Then we calculate the ISW effect over the resulting allowed ranges. 
In \S~(\ref{Results}) we will display planes in ${w_0,\ w_1}$~space with contours for the allowed 
$95.4\%\ (2\sigma)$ and $68.3\%\ (1\sigma)$ regions, and contours for the magnitude of the ISW effect.

\subsection{Initial conditions}
\label{sub:initial conditions}

We solve the system of Eqs.~(\ref{phiz})-(\ref{Gammaz}) for each of the three $w_{\rm DDE}$-parameterizations listed 
in Eqs.~(\ref{linder_param})-(\ref{RTparam}), and for a variety of parameter choices $(w_0, w_1)$. 
For each parameter choice, we evolve Eq.~(\ref{phiz}) between last scattering and now, 
i.e. a redshift range of $z=1100$ to $0$, eventually filling the $\{w_0,w_1\}$-plane.
Then we make another parameter choice and repeat the procedure, etc.

Several possibilities exist for choosing the initial ($z_i\sim 1100$) values of $S(z)$ and $\Gamma(z)$. 
However, in this work we focus on the adiabatic choice $S(z_i)=\Gamma(z_i)=0$. As pointed out in \cite{Bartolo:2003ad} 
(see also \cite{Wands:2000dp} and \cite{Malquarti:2002iu}), it is clear from Eqs.~(\ref{Sz})-({\ref{Gammaz}}) 
that on superhorizon scales ($k/aH\ll 1$), if perturbations are initially adiabatic, then the adiabaticity 
is preserved at all times. Most inflationary models (\cite{guthinflation}, see e.g. \cite{langlois} for a review) predict a scale invariant spectrum of adiabatic perturbations, and if the DDE perturbations were seeded by inflation, 
it is reasonable to assume that they were initially adiabatic. Initially adiabatic modes which are 
horizon-size today at z=0 have been superhorizon (and hence adiabatic) for most of the history of the Universe, 
and hence our choice $S(z_i)=\Gamma(z_i)=0$ seems well motivated. 
Of course, it is possible to solve Eqs.~(\ref{phiz})-(\ref{Gammaz}) 
to investigate the effect of choosing different initial values of the perturbations,
but we do not do so here.

For each parameter set,  we use the same initial values for $(\Phi_{k},\dot\Phi_k)$, namely, $(1,0)$. 
As a result of the linearity of Eq.~(\ref{phiz}), 
the initial choice for $\Phi_k$ is arbitrary; it cannot affect the physical results.  
Put another way, only the ratio of the initial and final $\Phi$'s is physical.
For $\dot\Phi_k$, the choice we have made is reasonable, in that $\Phi_k$ is practically frozen 
during the matter-dominated era when our initial conditions are set.

\section{Results}
\label{Results}

\subsection{Time evolution of variables}
\label{sub:time evolution}
\begin{figure}
	\epsfig{file=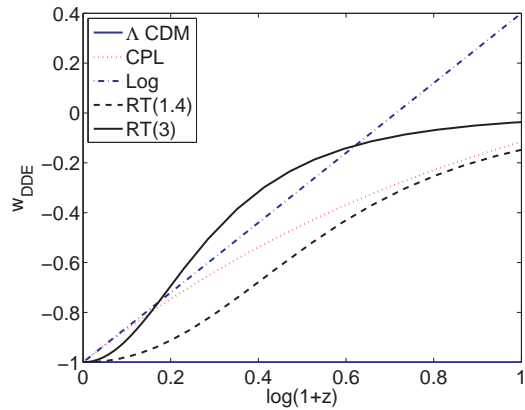,height=55mm}
	\caption
	{\label{wp}
$w_{\rm DDE}$ vs $z$ for the three DDE parameterizations with $(w_0,w_1)$=(-1,1.4),
and also for the RT parameterization with $(w_0,w_1)$=(-1,+3),
and for further comparison, the $w_\Lambda=-1$ $\Lambda$CDM case.
}
\end{figure}

\begin{figure}
	\epsfig{file=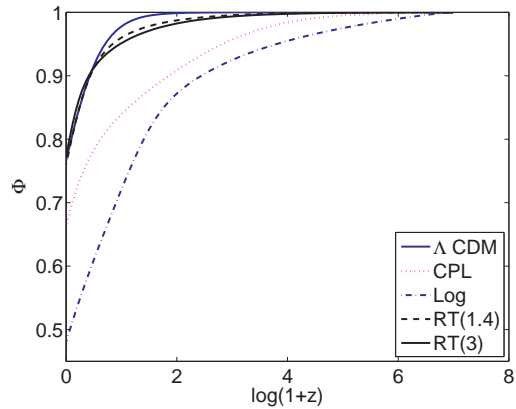,height=55mm}
	\caption
	{\label{Phi}
Potential $\Phi$ vs $z$ for the three scalar field DDE parameterizations with $(w_0,w_1)$=(-1,+1.4),
and also for the RT parameterization with $(w_0,w_1)$=(-1,+3),
and for $\Lambda$CDM.
}
\end{figure}

In Figs.~(\ref{wp}-\ref{Phi}), we show the $z$-dependence of the variables $w_{\rm DDE}$ and $\Phi$ 
for the three parameterizations described above. In each case, to visually demonstrate the effect, we choose the somewhat extreme parameter values $w_0=-1$ and $w_1=1.4$.For the RT parameterization, we also examine a more rapid transition having $w_0=-1$ and $w_1=3$. 
(In the RT parameterization, $w_{\rm DDE}$ is an even function of $w_1$, and so the sign of $w_1$ has no meaning.) 
All of these choices for $\{w_0,w_1\}$ (except for the last) are allowed at $95.4\%\ (2\sigma)$ confidence 
by the SNIa data, but not the CMB+BAO data.

Fig.~(\ref{wp}) shows the behavior of the scalar field DDE EoS $w(z)$ for our parameter choices for the different parameterizations. Note that the logarithmic parameterization gives physically unrealistic values of $w(z)$ for large $z$, but that is to be expected since this parameterization is unbounded for large $z$.

In Fig.~(\ref{Phi}) find that the late-time dominance of dark energy causes the gravitational 
potential to decrease as expected. In a $\Lambda$CDM cosmology, it is known that the gravitational potential changes 
by about ~25\% between last scattering and the present time \cite{huscranton}. This is evident from our plot as well. 
The other curves in Fig.~\ref{Phi} indicate that the decrease in $\Phi$ can be quite different from the $\Lambda$CDM scenario. 
The difference depends on the parameterization scheme chosen, and on the specific values of the parameters. 


\subsection{Comparison to observations}
\label{sub:observations}

We next address the question that motivated this work: 
for a range of ``allowed'' values of parameters in a given parameterization, 
how different can the ISW effect due to scalar field DDE perturbations be from that due to the standard $\Lambda$CDM? 
To answer this question, we compare the evolved $z=0$ value of $\Phi$ from a perturbed scalar field DDE scenario
to that obtained from evolution in $\Lambda$CDM cosmology (where $w=-1$ and $c^2=0$). 
We define a ``quality'' variable $Q$ to characterize the relative difference in the ISW effect for scalar field DDE models versus $\Lambda$CDM.
It is given by 
%
%

\begin{equation}
Q\equiv 1-\frac{\[\Delta T/T\]_{\rm ISW,DDE}}{\[\Delta T/T\]_{{\rm ISW},\Lambda{\rm CDM}}}
\end{equation}

For each of the three parameterizations listed in the previous section,
we evolve the metric perturbations $\Phi$ and $\Phi_\Lambda$ over the redshift range $z=1100$ to $0$. 
From these, we infer the $Q$ values.
We fill the $\{w_0,w_1\}$-plane with iso-$Q$ contours,
with contour values number-coded as shown in Table~\ref{CodingForQValues}. 
Notice that $Q<0$ ($Q>0$) means that the ISW effect is enhanced (suppressed) for scalar field DDE cosmology compared to the 
$\Lambda$CDM cosmology.  We will see that typically $Q<0$, which implies a larger ISW effect in DDE cosmology.

For each of the three parameterizations, 
we then use the SNIa standard candle data (ESSENCE+SNLS+HST from~\cite{Davis:2007na}),to construct a $\chi^2$ likelihood indicating which values of $(w_0,w_1)$ 
are allowed at the $68.3\%\ (1\sigma)$ and $95.4\%\ (2\sigma)$ confidence levels. 
The $\chi^2$ from SNIa is calculated as follows:
\be
\chi ^2 _{SN}  = \frac{{\sum\limits_{i = 1}^N {\left[ {\mu _{\text{obs} } \left( {z_i } \right) - \mu _{\rm th} \left( {z_i } \right)} \right]} ^2 }}{{\sigma^{2} _{\mu,i} }}
\ee
where $N=192$ is the number of SNIa data points. $\mu_{obs}$ is the  observed distance modulus, defined as the difference between the apparent and absolute magnitude of the supernova. The $\sigma_{\mu,i}$ are the errors in the observed distance moduli, arising from a variety of sources, and assumed to be gaussian and uncorrelated. The theoretical distance modulus $\mu_{\rm th}$  depends on the model parameters via the dimensionless luminosity distance $D_{L}(z)$:
\be
D_{L}\(z\)\equiv\left(1+z\right) \int^{z}_{0}dz'\frac{H_0}{H\left(z';\Omega_m(0),w_0,w_1\right)}
\ee
as follows:
\be
\mu_{\rm th}\left(z\right)=42.38-5\log_{10}h+5\log_{10}\[D_{L}\left(z\right)\]
\ee

From these results we construct constraint contours assuming $\Omega_m \(0\)=0.28$ (best-fit value from WMAP 5-year data \cite{Komatsu:2008hk}) and  marginalizing over the present day Hubble parameter $h$, 
following the techniques described in~\cite{perivol1,perivol2}.

\begin{table*}
	\centering
		\begin{tabular}{|l|l|}
		\hline
		\textbf{Number} & \textbf{Q-range}\\\hline
		$1$ & $Q<-80\%$ \\\hline
		$2$ & $Q=-80\% \text{ to } 60\%$ \\\hline
		$3$ & $Q=-60\% \text{ to } -40\%$ \\\hline
		$4$ & $Q=-40\% \text{ to }  -20\%$ \\\hline
		$5$ & $Q=-20\% \text{ to }  0\%$ \\\hline
		$6$ & $Q=0\% \text{ to }  20\%$ \\\hline
		\end{tabular}
	\caption{Number-coding for Q values.  Negative Q means that the ISW effect is enhanced with scalar field DDE relative to $\Lambda$CDM, 
	         whereas positive Q means that the ISW effect is relatively suppressed with DDE.}
	\label{CodingForQValues}
\end{table*}

Aside from the supernovae standard candle data, one can also derive physical constraints from observations relating to standard rulers, namely the CMB and the Baryon Acoustic Oscillations. Wang and Mukherjee showed that together with the baryon density parameter $\Omega_b h^2$, the ``CMB shift parameters'' \cite{Wang1,Wang2} defined as follows :
\be
R\equiv \sqrt{\Omega_m\(0\)}H_0 r\(z_*\),\,\quad l_{a}\equiv \pi r\(z_*\)/r_{s}\(z_*\)
\ee
can be used to set roughly model-independent constraints on dark energy models (see also \cite{Elgaroy:2007bv}). Here $r(z)$ is the comoving distance to redshift $z$ defined as:
\be
r(z)\equiv\int_{0}^{z}\frac{1}{H\(z\)}dz
\ee
$r_{s}\(z_*\)$ is the comoving sound horizon at decoupling (redshift $z_*$) given by
\be
r_{s}\(z_*\)=\int_{z_*}^{\infty}\frac{1}{H\(z\)\sqrt{3\(1+R_{b}/\(1+z\)    \)}}dz
\ee
The quantity $R_b$ is the photon-baryon energy-density ratio, and its value can be calculated as $R_b=31500 \Omega_{b} h^2 \(T_{CMB}/2.7K\)^{-4}$. The redshift at decoupling $z_*$ is given by the formulas in \cite{husugiyama}.

$R$ can be physically interpreted as a scaled distance to recombination, and $l_{a}$ is clearly the angular scale of the sound horizon at recombination. 

Following \cite{Komatsu:2008hk}, we compute the $\chi^2$ contribution of the CMB as  
\be 
\chi^{2}_{CMB}=\mathbf{V}_{\rm CMB}^{\mathbf{T}}\mathbf{C}_{\rm inv}\mathbf{V}_{\rm CMB}
\ee
Here $\mathbf{V}_{\rm CMB}\equiv\mathbf{P}-\mathbf{P}_{\rm data}$, where $\mathbf{P}$ is the vector $\(l_{a},R,z_{*}\)$ and the vector $\mathbf{P}_{\rm data}$ is formed from the WMAP $5$-year maximum likelihood values of these quantities \cite{Komatsu:2008hk}. The inverse covariance matrix $\mathbf{C}_{\rm inv}$ is provided in \cite{Komatsu:2008hk}. 

A second standard ruler is provided by measurements of the Baryon Acoustic Oscillation (BAO) peaks.  The measured quantity here is the ratio $r_{s}\(z_{*}\)/D_{V}\(z\)$, where the denominator is the so called ``volume distance'' defined in terms of the angular diameter distance $D_{A}\equiv r\(z\) /\(1+z\)$ as 
\be
D_{v}\(z\)\equiv\left[\frac{\(1+z\)^2 D_{A}^{2}(z) z }{H(z)}\right]^{1/3}
\ee
So far the BAO peak has been measured at two redshifts, $z=0.2$ and $z=0.35$ \cite{Seo:2005ys,Percival:2007yw}. The ratio of the two measurements of $D_{v}\(z\)$, i.e., $D_{v}\(.35\)/D_{v}\(.2\)=1.812\pm0.060$ \cite{Percival:2007yw}, can be used as a model-independent observational constraint. In this paper, we calculate the $\chi^2$ contribution of the BAO measurements as follows: 

\be 
\chi^{2}_{BAO}=\mathbf{V}_{\rm BAO}^{\mathbf{T}}\mathbf{C}_{\rm inv}\mathbf{V}_{\rm BAO}
\ee

The vector $\mathbf{V}_{\rm BAO}\equiv\mathbf{P}-\mathbf{P}_{\rm data}$, with $\mathbf{P}\equiv \( D_{v}\(0.32\),D_{v}\(0.2\) \) $ and $\mathbf{P}_{\rm data}\equiv\(0.1980, 0.1094\)$, the two measured BAO data points  \cite{Percival:2007yw}. The inverse covariance matrix is provided in \cite{Percival:2007yw}. 

\begin{figure}
	\epsfig{file=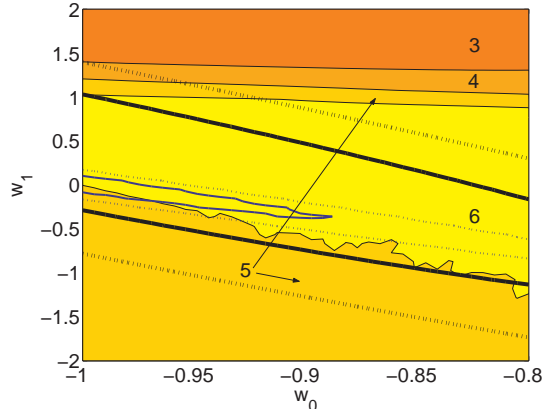,height=55mm}
	\caption
	{\label{linder}
$Q$ magnitudes in the $\{w_0,w_1\}$-plane for the CPL parameterization. 
The number-coding is as in table \ref{CodingForQValues}. 
The black (thick) lines show the $95.4\%$ (solid) and $68.3\%$ (dotted) contours from SNIa data are shown. The blue (thin) lines show the corresponding contours from the CMB+BAO data.
  }
\end{figure}

\begin{figure}
	\epsfig{file=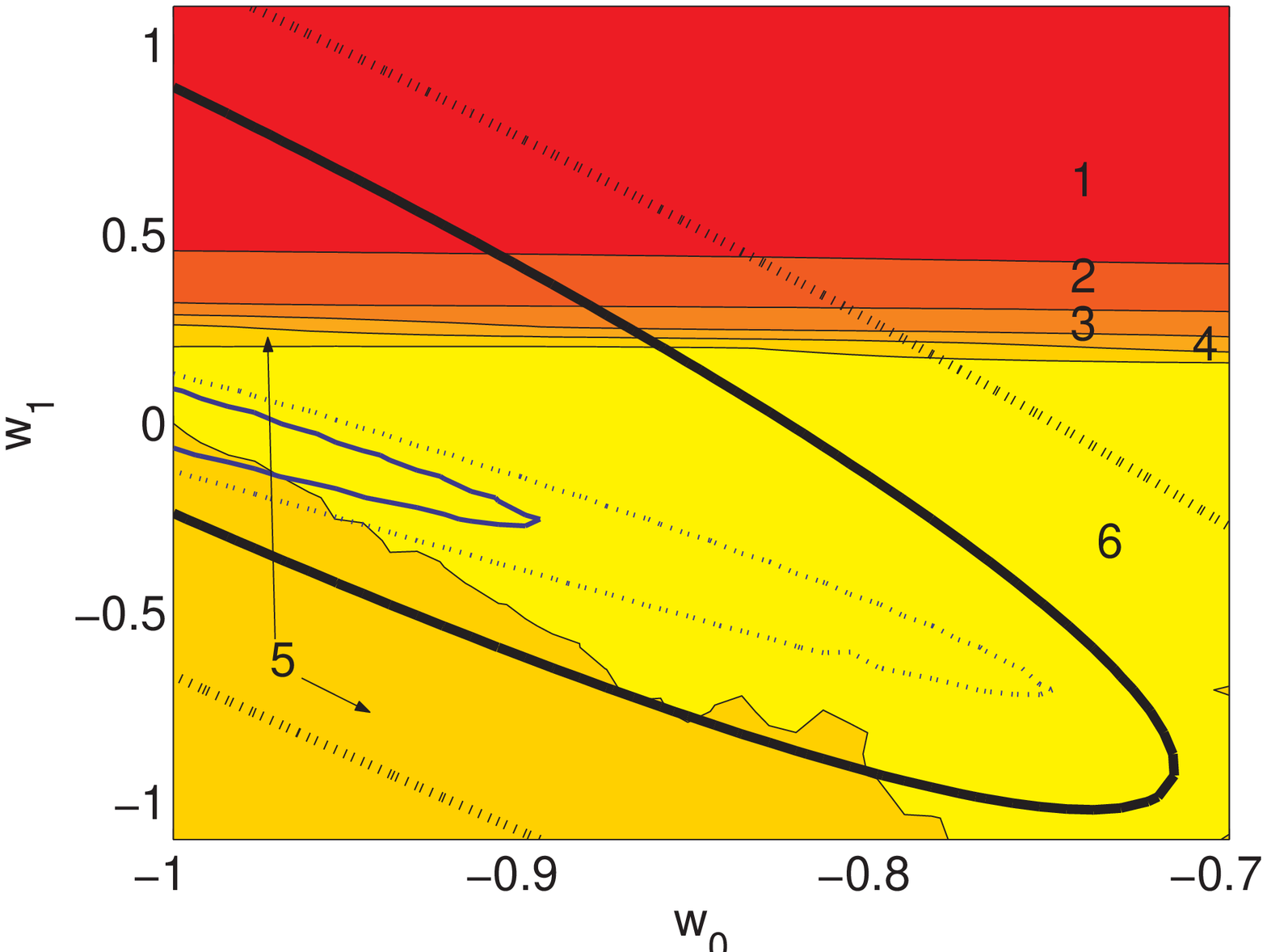,height=55mm}
	\caption
	{\label{log}
$Q$ magnitudes in the $\{w_0,w_1\}$-plane for the logarithmic parameterization. 
The number-coding is as in table \ref{CodingForQValues}. 
The black (thick) lines show the $95.4\%$ (solid) and $68.3\%$ (dotted) contours from SNIa data are shown. 
The blue (thin) lines show the corresponding contours from the CMB+BAO data.
  }
\end{figure}

\begin{figure}
	\epsfig{file=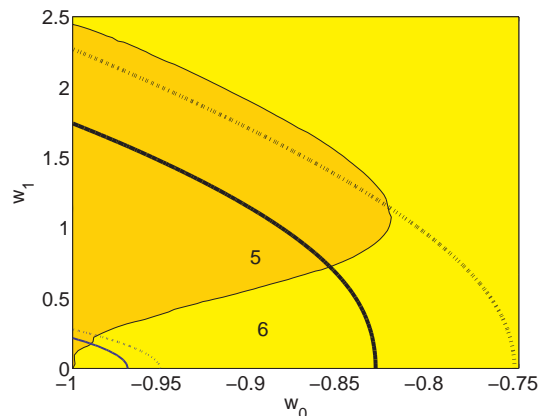,height=55mm}
	\caption
	{\label{kink}
$Q$ magnitudes in the $\{w_0,w_1\}$-plane for the RT parameterization. 
The number-coding is as in Table~\ref{CodingForQValues}. 
The RT parameterization is a symmetric function of $w_1$, and so the plot is symmetric about
$w_1=0$; we show only the positive half-plane $w_1>0$. 
The black (thick) lines show the $95.4\%$ (solid) and $68.3\%$ (dotted) contours from SNIa data are shown. 
The blue (thin) lines show the corresponding contours from the CMB+BAO data.
  }
\end{figure}

In Figs (\ref{linder}-\ref{kink}) we show the constraint-contours arising from both the standard candle (SNIa) and the standard ruler (CMB+BAO) data.

\subsection{Discussion}
\label{sub: Discussion}


The constraint contours allow one to determine the range of variation of Q within the regions allowed by the observational data. We note that the ISW effects are qualitatively similar for the CPL Fig.~(\ref{linder} and the logarithmic Fig.~(\ref{log} parameterizations. 
It is seen that at $2\sigma$, the $Q$ values range from about +20\% to -80\% (SN1a constraints) and from about  20\% to -20\% (CMB+BAO) constraints.From this one can reasonably conclude that our results are valid across a large class of scalar field DDE models in which the equation of state gradually evolves towards $-1$ at late times. 

The RT parameterization studies the scenario where the EoS reaches $-1$ in a sharp transition, and in this case the parameter $w_1$ determines the ``sharpness'' of the jump. Here the data indicate that while sharp ($w_1\gtrsim 2$) are allowed by the SN1a data, the $Q$ values range between  20\% to -20\%. 
 
%

%
The variation of the ISW signal from that expected from $\Lambda$CDM for certain ranges of parameters is highly interesting.
In some parts of parameter space for $w_{\rm DDE}(z)$,
the ISW signal is enhanced, by as much as 80\% for the logarithmic parameterization
On the other hand, the signal is suppressed in some of parameter space.

\section{Conclusions}
\label{Conclusions}

We have studied the impact of dynamical dark energy perturbations on the late-time ISW effect.  
We developed a  general gauge-invariant approach for evolving the growth of perturbations in the presence of a dynamical dark 
energy.  The approach is general in that it uses only a parameterization of the scalar field DDE EoS, 
rather than a specific Lagrangian. It also rigorously incorporates entropy perturbations in a consistent manner. As an interesting by-product, our formalism allows us to derive an explicit relationship (Eq.~\ref{cs_hat}) between the scalar field DDE rest frame sound speed $\hat{c}_{s,{\rm DDE}}^2$ and $w_{\rm DDE} (z)$, which shows that the evolution of these two quantities is linked, subject to initial conditions on the entropy perturbations $S$ and $\Gamma$. 

We found that if linear dark energy perturbations with adiabatic initial conditions evolve on horizon scales at low redshifts, 
they can enhance or suppress the ISW signal in the CMB, depending on the dark energy model used.
In the case of dark energy models in which the scalar field DDE EoS gradually evolves to $-1$ at late times, the suppression was as much as 10\%, 
and the enhancement was as much as 80\% for models allowed by the SNIa data at 95.4\% confidence. 
Models allowed by the standard ruler data (CMB+BAO) were found to enhance the ISW signal by  20\% or suppress it by 20\%. 
Our treatment can easily be extended to other parameterizations of the scalar field DDE component and/or different initial conditions 
on the DDE perturbations. 
%

As an interesting and necessary side issue, we have also used the SNIa data to place constraints on a simple 
2-parameter model in which the DDE EoS can evolve to $-1$ in a sharp transition. In this instance we have shown that while the data do support rapid transitions of the EoS parameter, the impact on the ISW effect is restricted to within $\pm 20\%$ of the $\Lambda$CDM effect . 

\begin{acknowledgements}
The authors are grateful to Niayesh Afshordi, Simon Dedeo, Dragan Huterer, Thomas Kephart, Irit Maor, Levon Pogosian, 
Robert Scherrer and the anonymous referee for useful discussions. SD acknowledges the hospitality of the 
Institute for Theoretical Science, University of Oregon, where part of this work was completed. 
\end{acknowledgements}

\end{document}